\documentclass[floatfix,aps,prd,showpacs,amsmath,nofootinbib,preprintnumbers,onecolumn]{revtex4}

\usepackage{graphicx}
\usepackage{bm}

\newcommand{\figurewidth}{4.in}

\def\half{{1\over 2}}

\def\half{{1\over 2}}
\def\({\left(}
\def\){\right)}
\def\[{\left[}
\def\]{\right]}

\def\e{\begin{equation}}
\def\q{\end{equation}}
\def\m{\begin{eqnarray}}
\def\n{\end{eqnarray}}

\begin{document}

\title{The (p,q) inflation model}

\author{Qing-Guo Huang}\email{huangqg@itp.ac.cn}
\affiliation{State Key Laboratory of Theoretical Physics, Institute of Theoretical Physics, 
Chinese Academy of Science, Beijing 100190, People's Republic of China}

\date{\today}

\begin{abstract}

In this paper we propose a new inflation model named (p,q) inflation model in which the inflaton potential contains both positive and negative powers of inflaton field in the polynomial form. We will show that such a simple model can easily  generate a large amplitude of tensor perturbation and a negative running of spectral index with large absolute value.

\end{abstract}

\pacs{}

\maketitle

%%%%%%%%%%%%%%%%%%%%%%%%%%%%%%%%%%%%%%%%
%%%%%%%%%%%%%%%%%%%%%%%%%%%%%%%%%%%%%%%

\section{Introduction}

Inflation \cite{Guth:1980zm} was proposed to solve the several puzzles, e.g. horizon problem, flatness problem and so on, in the hot big bang model. The simplest version is the so-called canonical single-field slow-roll inflation model \cite{Linde:1981mu,Albrecht:1982wi} in which the inflaton field slowly rolls down its potential. On the other hand, the quantum fluctuations during inflation can provide tiny primordial density perturbations which seed the anisotropies in the cosmic microwave background (CMB) and the large-scale structure of our universe. Since the Hubble parameter during inflation is roughly a constant, both the spectra of scalar and tensor perturbations are nearly scale-invariant. Usually the small scale dependence of scalar power spectrum is measured by the spectral index $n_s$ and then the amplitude of scalar power spectrum can be parametrized by 
\m
P_s=A_s \({k\over k_p}\)^{n_s-1}, 
\n
where $k_p$ is the pivot scale and $A_s$ is the amplitude at $k=k_p$. The size of tensor perturbation is characterized by the tensor-to-scalar ratio $r$ which is defined by $r\equiv P_t/P_s$, where $P_t$ is the amplitude of tensor power spectrum.

Combining with WMAP Polarization data \cite{Hinshaw:2012aka}, the Planck data \cite{Ade:2013zuv,Ade:2013uln} released in the early of 2013 imply $n_s=0.9603\pm 0.0073$ at the scale $k=0.05$ Mpc$^{-1}$. Recently BICEP2 \cite{Ade:2014xna} found an excess of B-mode power over the base lensed-$\Lambda$CDM expectation in the range $\ell\in [30,150]$. Cross correlating BICEP2 against 100 GHz maps from the BICEP1 experiment, the microwave mission by the polarized dust is disfavored at $1.7\sigma$. The observed B-mode power spectrum is well fitted by a lensed-$\Lambda$CDM+tensor model with 
\m
r=0.20_{-0.05}^{+0.07}. 
\n
Since the contaminant on the B-mode spectrum from the polarized dust gives a similar behavior in the multipoles around $\ell\sim 80$ \cite{Mortonson:2014bja,Flauger:2014qra}, it is still difficult to distinguish the signal of primordial gravitational waves from the dust. If the signal from BICEP2 is confirmed to be originated from the primordial gravitational waves by the upcoming data sets, it must be a breakthrough for the basic science. Because the primordial gravitational waves can also make a contribution to $C_{\ell}^{TT}$ on the low multipoles, it suppresses the contribution to $C_{\ell}^{TT}$ from scalar perturbation and hence $n_s>1$ for the scalar power spectrum is preferred on the large scales \cite{Cheng:2014hba}. In order to reconcile the different values of spectral index at different scales, the spectral index should be scale dependent: 
\m
n_s&=&1.0447_{-0.0297}^{+0.0295}, \\
{dn_s(k)\over d\ln k}&=& -0.0253\pm 0.0093, \label{exnrun}
\n 
at $k=0.002\ \hbox{Mpc}^{-1}$ in \cite{Cheng:2014bta}, where $\alpha_s\equiv dn_s/d\ln k$ is the running of spectral index. From the above results, we can estimate the spectral index at $k=0.05\ \hbox{Mpc}^{-1}$, namely $n_s(k=0.05\ \hbox{Mpc}^{-1})\simeq 1.0447-0.0253\times \ln(0.05/0.002)=0.9633$ which is consistent with the result from Planck at $k=0.05\ \hbox{Mpc}^{-1}$ \cite{Ade:2013zuv,Ade:2013uln}. See more discussion about the constraint on the running of spectral index in \cite{Ade:2014xna,Hu:2014aua}.

Usually the canonical single-field slow-roll inflation model predicts $|\alpha_s|\lesssim {\cal O}((n_s-1)^2)\sim 10^{-3}$ which is much smaller than that in Eq.~(\ref{exnrun}). In the literatures, there two possible ways to achieve a negative running of spectral index with large absolute value: one is the space-time noncommutative inflation \cite{Huang:2003zp,Huang:2003hw,Huang:2003fw}, the other is inflation with modulations \cite{Feng:2003mk,Kobayashi:2010pz,Czerny:2014wua}. In this letter we will propose a new inflation model, called (p,q) inflation, which can also easily achieve a large negative running of spectral index. 
This letter will be organized as follows. The introduction and qualitative discussion on (p,q) inflation model will be given in Sec.~2. The numerical predictions of (2,1) inflation model show up in Sec.~3. Discussion is included in Sec.~4. The more accurate formula for the runnings of spectral indexes of both scalar and tensor perturbations are given in the Appendix \ref{ap}.

\section{The $(p,q)$ inflation model}

From the dynamics of inflaton field during inflation, one can easily find that the evolution of inflaton field is related to the tensor-to-scalar ratio by 
\m
{|\Delta \phi|}\simeq {1\over \sqrt{8}} \int_0^N \sqrt{r} dN',
\n
where $N$ is the number of e-folds before the end of inflation. In this letter we work in the unit of $8\pi G=1$. The above relation is called the Lyth bound \cite{Lyth:1996im}. 
If the signal from BICEP2 stands after further cosmological observations and turns out to be primordial, the Lyth bound implies that a large field inflation model should be called for.

It is well-known that the inflation model with potential $V(\phi)\propto \phi^n$ 
is a typical large field inflation model, where $n$ can be positive \cite{Linde:1983gd} or negative \cite{Barrow:1990vx}. Generically the potential can take the form 
\m
V(\phi)=\cdots + c_{-1}{1\over \phi}+c_0+c_1\phi+c_2\phi^2+\cdots. 
\label{gpt}
\n
When the inflaton field is not so large, the terms with negative powers are dominant and the slow-roll parameter $\epsilon$ decreases with time, and hence the inflation cannot naturally exist if these terms are always dominant. However, the terms with positive powers become important when the inflaton field $\phi$ goes to the region far away from $\phi=0$. One may expect that the inflation ended and our universe was re-heated around the local minimum of the potential where $c_0$ is set by the condition that the potential equals zero at its local minimum.

In this letter we only consider a special form of the potential in Eq.~(\ref{gpt}), namely 
\m
V(\phi)=\lambda \[ s  {\phi_*^{p+q}\over \phi^q}+(\phi_*-\phi)^p \]+V_c,
\label{pt}
\n
where both $p$ and $q$ are positive, $s$ is a positive dimensionless parameter and $V_c$ is fixed by the condition that the minimum of potential is equal to zero. Here we also suppose $(\phi_*-\phi)^p=(\phi-\phi_*)^p$. The inflation model driven by the potential in Eq.~(\ref{pt}) is called (p,q) inflation model. For $s\ll 1$, the local minimum of the potential is located at 
\m
\phi=\phi_m\simeq \phi_*+\({q\over p} s\)^{1\over p-1} \phi_*, 
\n
and the potential in Eq.~(\ref{pt}) can be roughly written by 
\m
V(\phi)\simeq \lambda \[ s  \({\phi_*^{p+q}\over \phi^q}-\phi_*^p \)+(\phi_*-\phi)^p \], 
\label{ptp}
\n
up to the order of ${\cal O}(s^{1+1/(p-1)})$. The term with negative power is roughly the same as that with positive power when 
\m
\phi=\phi_c\simeq s^{1/q} \phi_*, 
\n 
and the contribution to $dV(\phi)/d\phi$ from the term with negative power is comparable to that from the term with positive power when 
\m
\phi= \phi_T\simeq \({q\over p}s\)^{1\over q+1} \phi_*. 
\n
Therefore, roughly speaking, the term $(\phi_*-\phi)^p$ is dominant when $\phi\gg \phi_T$ and the number of e-folds from $\phi=\phi_T$ to the end of inflation is roughly given by $N_T\simeq (\phi_*^2-\phi_T^2)/(2p)$. If $N_T$ is much larger than the number of e-folds corresponding to the CMB scales, e.g. $N_{\rm CMB}\simeq 60$, or equivalently $\phi_*\gg \sqrt{2pN_{\rm CMB}}$, the relevant predictions of (p,q) inflation model are the same as those of the chaotic inflation with potential $V(\phi)\sim \phi^p$. On the other hand, the term of $s\phi_*^{p+q}/\phi^q$ becomes dominant when $\phi<\phi_c$ and the spectral index is $n_s\sim 1+{2-q\over 2(N-N_T)}$ which is larger than one if $0<q<2$. So the (p,q) inflation model can possibly achieve a scalar power spectrum which is blue tilted on the very large scales and red tilted on the small scales.

For simplicity, we introduce a new variable which is related to $\phi$ by 
\m
\varphi\equiv \phi/\phi_*, 
\n
and then the potential in Eq.~(\ref{ptp}) reads 
\m
V_\varphi=s  (\varphi^{-q}-1)+(1-\varphi)^p,  
\n
where 
\m
V_\varphi\equiv V(\phi)/\lambda \phi_*^p. 
\n
Now the slow-roll parameters can be written by  
\m
\epsilon&=& {1\over 2\phi_*^2} \({V_\varphi'\over V_\varphi}\)^2, \\
\eta&=&{1\over \phi_*^2}{V_\varphi''\over V_\varphi}. 
\n
The number of e-folds before the end of inflation is given by  
\m
N\simeq -\phi_*^2 \int_{\varphi_N}^1 {V_\varphi\over V_\varphi'}d\varphi =\phi_*^2 \int_{\varphi_N}^1 {s(\varphi^{-q}-1)+(1-\varphi)^p \over sq \varphi^{-q-1}+p(1-\varphi)^{p-1}} d\varphi. 
\n
In the limit of $s\rightarrow 0$, the (p,q) inflation model reduces to the chaotic inflation model with potential $V(\phi)\sim \phi^p$ as well. In general, we cannot get the analytical formula for the above integration. For an instance, the numerical predictions of (2,1) inflation model will be discussed in the next section.

\section{Numerical predictions of (2,1) inflation model}

\begin{figure}[hts]
\centerline{\includegraphics[width=\figurewidth]{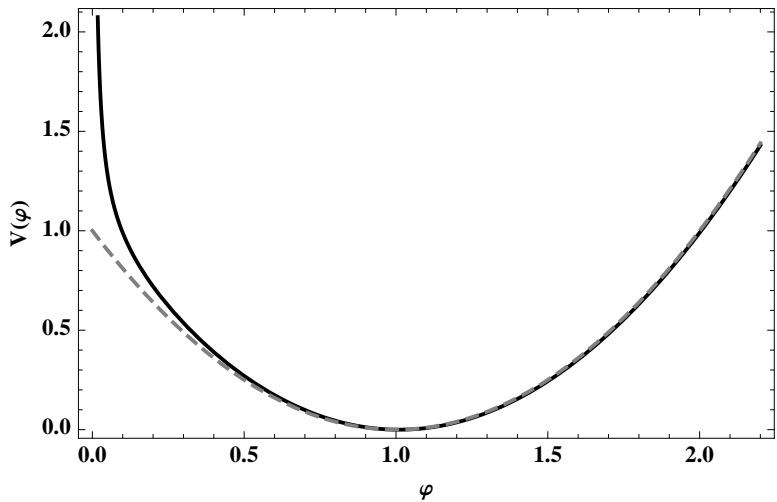}}
\caption{The potential of inflaton field $\varphi$ in the (2,1) inflation model. The grey dashed curve corresponds to the case with $s=0$.}
\label{fig:pt}
\end{figure}

\begin{figure}[hts]
\centerline{\includegraphics[width=\figurewidth]{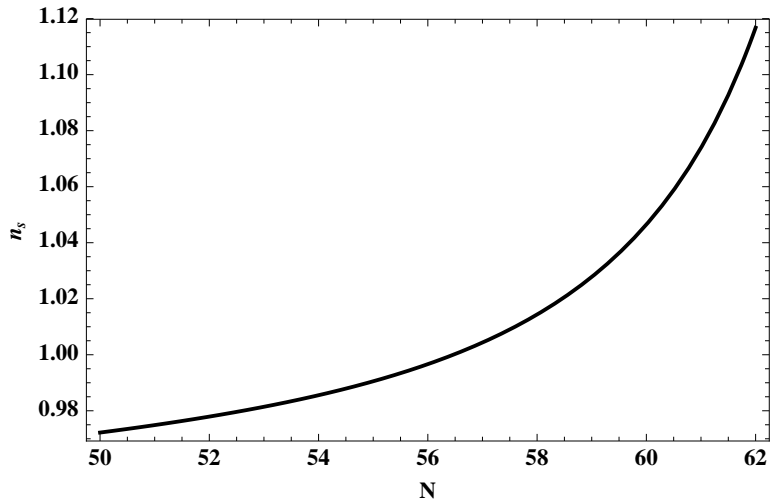}}
\caption{The spectral index $n_s$ varies with the number of e-folds $N$ before the end of inflation, where $\phi_*\simeq 17.7$ and $s=0.0083$.}
\label{fig:nsn}
\end{figure}

In this section let's consider a simple example, namely (2,1) inflation model in which the potential takes the form 
\m
V(\phi)=\lambda \[s  \( {\phi_*^3\over \phi}-\phi_*^2\)+(\phi_*-\phi)^2\], 
\n
and then 
\m
V_\varphi\simeq s  (\varphi^{-1}-1)+(1-\varphi)^2, 
\n
where $s \ll 1$. The shape of inflaton potential in (2,1) inflation model is illustrated in Fig.~\ref{fig:pt}. 
Now the slow-roll parameters are 
\m
\epsilon&\simeq&{1\over 2\phi_*^2} \({s \varphi^{-2}+2(1-\varphi)\over s  (\varphi^{-1}-1)+(1-\varphi)^2}\)^2, \\
\eta&\simeq& {1\over \phi_*^2} {2+2s \varphi^{-3}\over s  (\varphi^{-1}-1)+(1-\varphi)^2}, \\
\xi&\simeq&  {1\over \phi_*^4} {6s\varphi^{-4}(s \varphi^{-2}+2(1-\varphi)) \over [s  (\varphi^{-1}-1)+(1-\varphi)^2]^2}, \\
\sigma&\simeq& {1\over \phi_*^6} {24s\varphi^{-5} [s \varphi^{-2}+2(1-\varphi)]^2 \over [s  (\varphi^{-1}-1)+(1-\varphi)^2]^3}, 
\n
where the value of field $\varphi$ is related to the number of e-folds before the end of inflation by 
\m
N\simeq\phi_*^2 \int_{\varphi_N}^1 {s  (\varphi^{-1}-1)+(1-\varphi)^2 \over s \varphi^{-2}+2(1-\varphi)} d\varphi. 
\n
The amplitude of the scalar power spectrum is given by 
\m
P_s={\lambda \phi_*^4\over 12\pi^2} {(s  (\varphi^{-1}-1)+(1-\varphi)^2)^3\over (s \varphi^{-2}+2(1-\varphi))^2}. 
\n
In this model, there are three parameters, namely $\lambda$, $s$ and $\phi_*$. But there are four observables, i.e. $P_s$, $r$, $n_s$ and $\alpha_s$. The more accurate formula for these four observables are given in the appendix \ref{ap}. For example, considering that the tensor-to-scalar ratio and the spectral index are  $r=0.22$ and $n_s=1.0447$ respectively \cite{Cheng:2014bta} at the pivot scale $k_p=0.002$ Mpc$^{-1}$ assuming to correspond to $N=60$, we get $\phi_*\simeq 17.7$ and $s=0.0083$, and then it predicts $\alpha_s=-0.0226$ which is nicely consistent with the constraint from data in Eq.~(\ref{exnrun}).  In order to see how the spectral index runs with the perturbation scales, or equivalently the number of e-folds before the end of inflation, we plot the spectral index $n_s$ as a function of $N$ in Fig.~\ref{fig:nsn}.
From this figure, we clearly see that the spectral index in our model can run from $n_s>1$ to $n_s<1$ in a few number of e-folds. 
Using the normalization $P_{s,obs}\simeq 2.2\times 10^{-9}$ \cite{Cheng:2014bta}, we find $\lambda\simeq 2.54 \times 10^{-11}$.

\begin{figure}[hts]
\centerline{\includegraphics[width=2.3in]{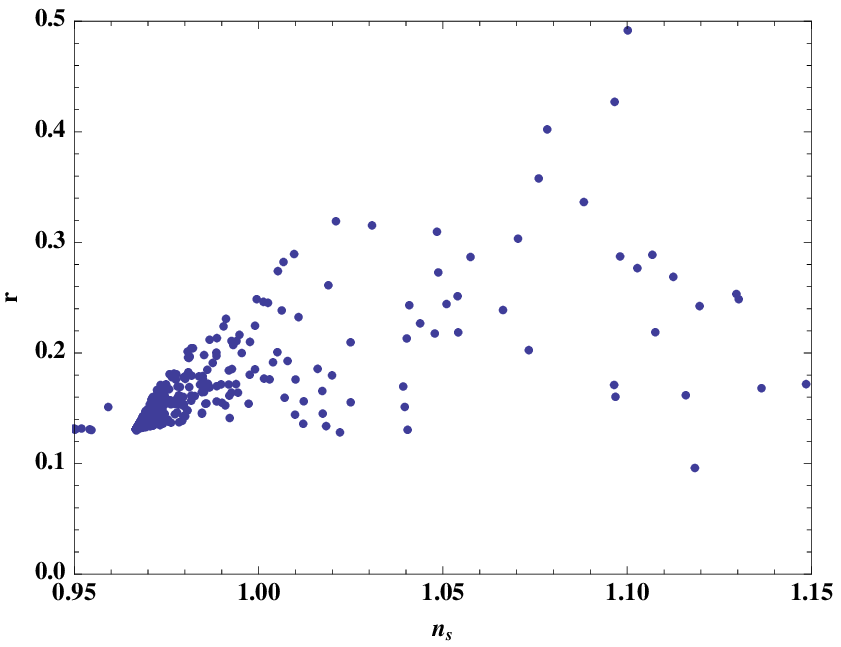}\quad \includegraphics[width=2.37in]{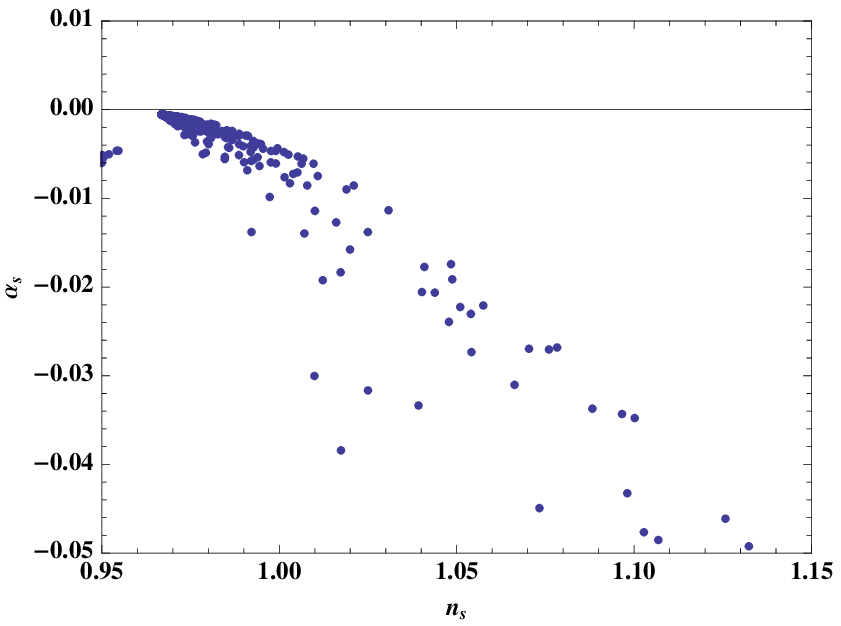}\quad \includegraphics[width=2.3in]{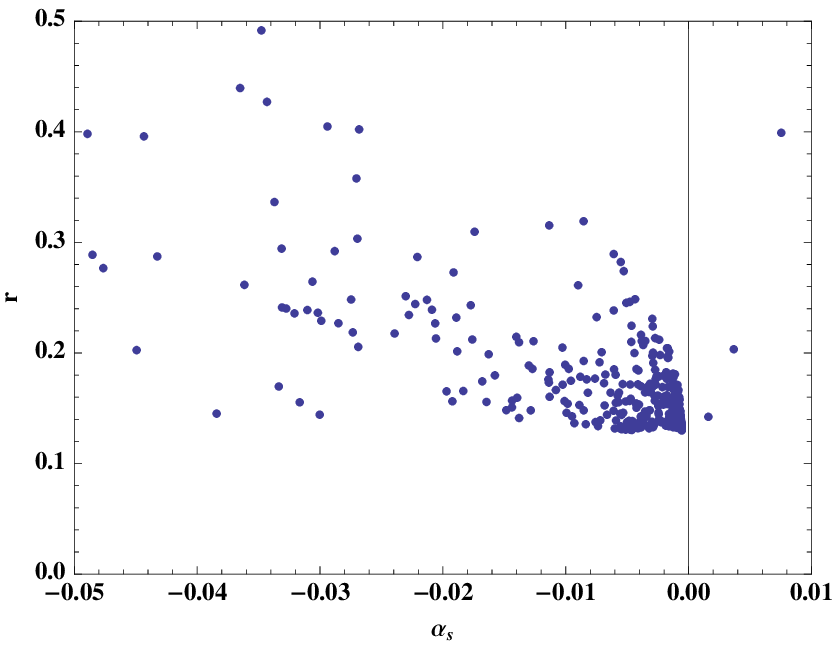}}
\caption{The predictions of (2,1) inflation model. }
\label{fig:preds}
\end{figure}

More generally, we take the parameters $\phi_*$ and $s$ as two free parameters and figure out the predictions of (2,1) inflation model. Considering $\phi_*\in [15,25]$ and $s\in [10^{-4}, 10^{-1}]$, we plot the numerical predictions of (2,1) inflation model in Fig.~\ref{fig:preds}. 
From Fig.~\ref{fig:preds}, we see that the (2,1) inflation model can easily generate a negative running of spectral index with large absolute value.

\section{Discussion}

The chaotic inflation model in which the potential takes the form $V(\phi)\sim \phi^p$ with $p>0$ predicts a red tilted scalar power spectrum with negligible running of spectral index. In order to achieve a blue tilted scalar power spectrum on the very large scales and a negative running of spectral index with large absolute value, we introduce an extra term with negative power in this letter. Usually it is hard to understand the term with negative power from the viewpoint of field theory. In order to understand such kind of term, we believe that a deeper insight on the field theory is needed. However, the inverse power term typically exists in string theory. For example, the potential of inflaton field in brane inflation \cite{Dvali:1998pa} takes the form $V(\phi)=V_0(1-\mu^4/\phi^4)$. The term of $-{1/\phi^4}$ can be easily understood from the viewpoint of gravity in the bulk. Unfortunately, there is a minus sign for the inverse power term which is originated from the fact that the gravity between brane and anti-brane is attractive. How to understand a positive term with negative power is still an open question. We expect that it may shed a light on the new fundamental physics.

\vspace{5mm}
\noindent {\bf Acknowledgments}

This work is supported by the project of Knowledge Innovation Program of Chinese Academy of Science and grants from NSFC (grant NO. 10821504, 11322545 and 11335012).

\appendix

\section{Formula for the tensor-to-scalar ratio, spectral indexes and their runnings of both scalar and tensor perturbations}
\label{ap}

An accurate analytic calculation of the spectrum of cosmological perturbations generated during inflation shows up in \cite{Stewart:1993bc,Huang:2006yt}. The amplitudes of scalar and tensor perturbation power spectra are given by 
\m
P_s&\simeq&\[1+{25-9c\over 6}\epsilon-{13-3c\over 6}\eta\] {V\over 24\pi^2\epsilon}, \\
P_t&\simeq& \[1-{1+3c\over 6}\epsilon\] {V\over 3\pi^2/2}, 
\n
where $c\simeq 0.08145$. 
Therefore the tensor-to-scalar ratio $r$, the spectral index of tensor perturbation $n_t$ and its running $\alpha_t$, the spectral index of scalar perturbation $n_s$ and its running $\alpha_s$ and the running of running $\beta_s\equiv d^2n_s/d\ln k^2$ can be written by  
\m
r&\equiv&{P_t\over P_s}\simeq 16\epsilon\[1-{13-3c\over 6}(2\epsilon-\eta)\],\\ 
n_t&\equiv&{d\ln P_t\over d\ln k}\simeq -2\epsilon-{2(2+3c)\over 3}\epsilon^2-{1-3c\over 3}\epsilon\eta,\\ 
\alpha_t&\equiv& {dn_t\over d\ln k}\simeq -8\epsilon^2+4\epsilon\eta-{8(5+6c)\over 3}\epsilon^3+2(1+7c)\epsilon^2\eta+2(1-c)\epsilon\eta^2+{1-3c\over 3}\epsilon\xi, \\
n_s&\equiv& 1+{d\ln P_s\over d\ln k}\simeq 1-6\epsilon+2\eta+{2(22-9c)\over 3}\epsilon^2-2(7-2c)\epsilon\eta+{2\over 3}\eta^2+{13-3c\over 6}\xi, \\
\alpha_s&\equiv&{dn_s\over d\ln k}\simeq -24\epsilon^2+16\epsilon\eta-2\xi+{8(41-18c)\over 3}\epsilon^3-{4(109-36c)\over 3}\epsilon^2\eta+4(9-2c)\epsilon\eta^2 \nonumber \\
&& \quad\quad\quad\quad +2(11-3c)\epsilon\xi-{25-3c\over 6}\eta\xi-{13-3c\over 6}\sigma, \\
\beta_s&\equiv& {dn_s^2\over d\ln k^2}\simeq -192\epsilon^3+192 \epsilon^2\eta-32 \epsilon\eta^2-24\epsilon\xi+2\eta \xi+2\sigma \nonumber \\ 
&& \quad\quad\quad\quad +96(13-6c)\epsilon^4-{8(791-288c)\over 3}\epsilon^3 \eta+{16(173-48c)\over 3}\epsilon^2 \eta^2 - {8(31-6c)\over 3}\epsilon \eta^3 \nonumber \\ 
&& \quad\quad\quad\quad +{4(235-72c)\over 3}\epsilon^2\xi - {511-111c \over 3}\epsilon \eta \xi+{29-3c\over 6}\eta^2\xi+{25-3c\over 6}\xi^2 \nonumber \\ 
&& \quad\quad\quad\quad -{103-27c\over 3}\epsilon \sigma + {55-9c\over 6}\eta\sigma+{13-3c\over 6} \vartheta, 
\n
respectively, where  
\m
\epsilon&\equiv& \half \({V'(\phi)\over V(\phi)}\)^2, \\
\eta&\equiv&{V''(\phi)\over V(\phi)}, \\
\xi&\equiv&{V'(\phi)V'''(\phi)\over V^2(\phi)}, \\
\sigma&\equiv& {V'^2(\phi)V''''(\phi)\over V^3(\phi)}, \\
\vartheta&\equiv& {V'^3(\phi)V^{(5)}(\phi)\over V^4(\phi)}.
\n

%%%%%%%%%%%%%%%%%%%%%%%%%%%%%%%%%%%%%%%%
%%%%%%%%%%%%%%%%%%%%%%%%%%%%%%%%%%%%%%%%

%%%%%%%%%%%%%%%%%%%%%%%%%%%%%%%%%%%%%%%%
%%%%%%%%%%%%%%%%%%%%%%%%%%%%%%%%%%%%%%%%

%%%%%%%%%%%%%%%%%%%%%%%%%%%%%%%%%%%%%%%%
%%%%%%%%%%%%%%%%%%%%%%%%%%%%%%%%%%%%%%%%
\end{document}